\documentclass{aa}  

\usepackage{color}

\usepackage{graphicx}
\usepackage{lipsum}  
\usepackage{mathtools}
\usepackage{siunitx}
\usepackage[bottom]{footmisc}
\usepackage{tablefootnote}
\usepackage[flushleft]{threeparttable}
\usepackage{adjustbox}
\usepackage{subcaption}
\usepackage{float}

\usepackage{txfonts}
\begin{document}

   \title{Is the correlation between the bulge-to-total stellar mass ratio and the number of dwarf galaxies in tension with $\Lambda$CDM?}
   \author{Oliver Müller
          \inst{1}
                    \and
          Ethan Crosby\inst{2}
          }

\titlerunning{The bulge-to-total stellar mass to number of satellites  relation}
   \institute{Institute of Physics, Laboratory of Astrophysics, Ecole Polytechnique F\'ed\'erale de Lausanne (EPFL), 1290 Sauverny, Switzerland\\
              \email{oliver.muller@epfl.ch}
   \and
             Research School of Astronomy and Astrophysics, Australian National University, Canberra, ACT 2611, Australia
             }
   \date{Received tba; accepted tbd}

 
  \abstract
   {
Previous results suggest that there exists a correlation between the size of the bulge of a galaxy and the number of its dwarf galaxy satellites. This was found to be inconsistent with the standard model of cosmology based on comparisons to semi-analytical dark matter-only simulations, where no such correlation was found. In this work, we extend these studies using the volume-complete ELVES dwarf galaxy catalog, which  increases the number of systems compared to previous work by a factor of four. For each giant galaxy we compile the bulge-to-total baryonic mass (B/T) ratio and put it as a function of the number of dwarf galaxies surrounding them within 250\,kpc (N$_{250}$). For the 29 galaxy systems in the ELVES catalog, we find a linear relation between B/T and N$_{250}$ which is consistent with previous data. However, for a given stellar mass of the host galaxy this relation is mainly driven by their morphology, where early-type galaxies have a larger B/T ratio and a larger N$_{250}$ than late type galaxies. 
By investigating spiral galaxies in Illustris-TNG100, we tested whether the inclusion of baryons in the simulations will change the results based on Millennium-II. Contrary to dark matter-only simulations, we do find a correlation between B/T and  N$_{250}$, indicating that the standard model of cosmology does predict a correlation. The empirical relation between the number of satellites and the bulge to total stellar mass is therefore not {necessarily in} tension with $\Lambda$CDM. 
}


   \maketitle
%

\section{Introduction}

\citet{2005AJ....129..178K} noted a peculiar connection between the abundance of dwarf galaxy satellites and the bulges of the host galaxy in nearby galaxy groups: the larger the central bulge of a galaxy, the more dwarf galaxies surrounded this galaxy. This is most evident in our own Local group, where the Andromeda galaxy hosts a more massive bulge \citep{2017ApJ...837L...8B} as well as more dwarf galaxies \citep{2012AJ....144....4M}. This may not seem surprising, because the Andromeda galaxy is likely twice as massive as the Milky Way \citep{2022MNRAS.513.2385C,2022arXiv221115928P}. It is well known that the number of dwarf galaxies is directly correlated to the mass of the central dark matter halo. However, for similar massive dark matter halos a similar number of dwarf galaxies is expected (within a scatter, e.g. {\citealt{MuellerTRGB2019,2020MNRAS.491.1471S,2021ApJ...908..109C}}). Therefore it is intriguing that \citet{2019ApJ...870...50J} found that there is  no correlation between the bulge size of the central galaxy and the number of dwarf galaxies for a given mass in cosmological dark matter-only simulations with semi-analytical galaxy formation models.

Using classifications of bulge sizes from the Galaxy Zoo citizen project \citep{2013MNRAS.435.2835W} and a catalog of tidal dwarf galaxies  \citep{2012MNRAS.419...70K}, \citet{2016ApJ...817...75L} found a relation between the number of tidal dwarf galaxies and the size of the bulge for same-mass host galaxies. They argue that such a relation is not expected in the standard model of cosmology (as later {indicated} by \citealt{2019ApJ...870...50J}), but very well in alternative gravity scenarios like modified Newtonian dynamics (MOND, \citealt{Milgrom1983a}), which, however, still needs a more quantitative analysis (some simulations of the formation of tidal dwarfs in MOND were e.g. done in {\citealt{2008ASPC..396..259T,2018A&A...614A..59B,2018MNRAS.473.4033B}}).  {While in a MOND context tidal dwarf galaxies and dwarf galaxies may have the same origin\footnote{{Currently, the formation of dwarf galaxies in MOND has only been simulated through tidal interactions, but this is due to the construction of the simulation. To date, no MOND cosmology exists, therefore the nature and origin of dwarf galaxies in MOND is unknown. }}}, this is, they {may} form in tidal interactions, they are fundamentally different within $\Lambda$ Cold Dark Matter ($\Lambda$CDM), representing different generations of galaxies. Predicting the number of tidal dwarf galaxies in the standard $\Lambda$CDM model is difficult, because it needs high-resolution baryonic simulations in cosmological contexts \citep{2018MNRAS.474..580P}. The different origins of these dwarf galaxies in $\Lambda$CDM means that tidal dwarf galaxies can not be used as tracers for the overall dwarf galaxy population (including primordial and tidal dwarf galaxies). And therefore,  in standard cosmology, from tidal dwarf galaxies no conclusions based on the abundance of satellite galaxies and the bulge mass of the host can be drawn. There is no way around than to study the primordial dwarf galaxy populations in nearby groups, which is a difficult task due to their faintness and low-surface brightness {\citep[e.g., ][]{2017ApJ...848...19P,2020MNRAS.491.1901H,2023arXiv230513966C}}.

Recent efforts expanded our knowledge of satellite systems in the nearby universe. Multiple teams searched for dwarf galaxies using different instruments, such as the Dark Energy Camera \citep{2015A&A...583A..79M,2017MNRAS.469.3444T,2020A&A...644A..91M}, the MegaCam \citep{2009AJ....137.3009C,2016ApJ...823...19C,2020MNRAS.491.1901H}, the OmegaCam \citep{2019A&A...625A.143V,2022A&A...665A.105L}, the KMTNet \citep{2020ApJ...891...18B,2023arXiv230316849J} the Hyper Suprime Cam \citep{2018ApJ...863..152S,2019ApJ...884..128O,2022ApJ...937L...3B,2023MNRAS.521.4009C}, the Dragonfly array \citep{2014ApJ...787L..37M,2018ApJ...868...96C}, or even with small amateur telescopes \citep{2015AstBu..70..379K,2016A&A...588A..89J,2017A&A...603A..18H}. These efforts uncovered a plethora of hitherto undetected dwarf galaxies. 

\citet{2020ApJ...891..144C} exploited  archival MegaCam data to survey ten nearby galaxy groups. \citet{2020MNRAS.493L..44J} used this data together with literature values to study the relation between the number of dwarf galaxy satellites and the bulge-to-total baryonic mass ratio (B/T). Because B/T is a mass ratio, it is not dependent on the total baryonic mass of the galaxy but rather gives an estimate about how prominent the bulge is. Using seven host galaxies, \citet{2020MNRAS.493L..44J} found a linear relation between B/T and the number of satellites, which seems to be at odds with cosmological $\Lambda$CDM predictions as shown in \citet{2019ApJ...870...50J}, adding to the list of other known small-scale problems \citep[e.g., ][]{2011MNRAS.415L..40B,2012MNRAS.423.1109P,2017MNRAS.468L..41C,Muller2021b,2022Natur.605..435V}, see also the reviews by \citet{2010A&A...523A..32K}, \citet{2017ARA&A..55..343B}, or \citet{2022NatAs...6..897S}.

In this paper, we test the relation between B/T and the number of satellites by  increasing the number of studied systems by a factor of four compared to previous work. In Section \ref{data} we present the data used in this work, in Section \ref{relation} we study the B/T  to  N$_{sat}$  relation, in Section \ref{simulations} we compare the data to Illustris-TNG100, and in Section \ref{summary} we give a summary and conclusion.

\section{Data}
\label{data}

We use the dwarf galaxy satellite catalog from \citet{2022ApJ...933...47C} -- Exploration of Local VolumE Satellites (ELVES) -- which is a combination of their own survey as well as a compilation of archival data (see the references under Table\,\ref{tab:sample}). ELVES represents a volume complete dwarf galaxy catalog within 10\,Mpc. {The detection completeness is M$_V$=-9\,mag based on artificial galaxy experiments.} In total they provide data for 30 galaxies. We exclude the Milky way in our analysis, because the detection and completeness estimation is quite different than for external systems, leaving us a sample of 29 host galaxies.
\citet{2022ApJ...933...47C} provide the positions of all dwarf galaxy satellites, as well as the survey {footprint} limit (r$_{\rm cover}$). The average radial coverage is 266\,kpc, which is close to the radial selection of dwarf galaxies of 250\,kpc used by  \citet{2020MNRAS.493L..44J}. We adpot the same cut {of 250\,kpc} here. The systems contain between 3 and 66 dwarf galaxies. If we select only dwarf galaxies within 250\,kpc, the abundance ranges between 3 to 40. In Table\,\ref{tab:sample} we provide the numbers of satellites within 250\,kpc (N$_{250}$), but because not all systems cover this radial extent, we additionally apply corrections. This is the case for NGC\,4258, NGC\,4565, M104, and NGC\,5194 with a radial coverage of 150\,kpc and NGC\,4631, Cen\,A, and NGC\,6744 with 200\,kpc.  To apply a correction we calculate the average difference between the number of dwarf galaxies for systems with a coverage larger than 250\,kpc. We take two differences, one between 250\,kpc and 150\,kpc and one between 250\,kpc and 200\,kpc. The average difference is 5 and 3 dwarfs, respectively. By adding these to N$_{250}$ of the incomplete systems (NGC\,4258, NGC\,4565, M104, NGC\,5194, Cen\,A, and NGC\,6744) we get a corrected number of dwarf galaxies within 250\,kpc (N$_{250}^{\rm cor}$).

We did not find an estimation of B/T in the literature for all galaxies in our sample. Where we did not find the number directly, we searched for the bulge mass and disk mass or luminosity individually and calculated the B/T ratio ourselves. This was the case for NGC\,628, NGC\,1023, and NGC\,4826 where we converted the $J$ band photometry of the bulge and the whole galaxy into stellar masses and adopted a conservative error of 30\% on the B/T value. For NGC\,1291 we found the spheroidal to total luminosity ratio of 0.57 \citep{1975ApJS...29..193D}, without an error. We take this value as B/T and take an uncertainty of 30\% on B/T. For NGC\,1808, NGC\,3344, NGC\,4826, NGC\,5055, NGC\,5194, NGC\,5457, and NGC\,6744 the masses are given without uncertainties, so we assume an error of 15\% on both the bulge and total mass. For NGC\,4517 a B/T value of 0.02 is given with no uncertainty, which we set to 0.02 (i.e. consistent with no bulge). For M104, a B/T value of 0.77 \citep{2012MNRAS.423..877G} is given without uncertainty. Because two  other studies give a value of 0.77 and 0.73 for M104, respectively \citep{2006ApJ...645..134B,2011ApJ...739...21J}, we take the mean and standard deviation of these three values as our B/T and the corresponding uncertainty.
For four galaxies, no B/T ratio can be calculated, because they are either elliptical galaxies, or have no visible bulge. For the former (NGC\,3379), we assign a B/T ratio of 1, for the latter (NGC\,3556, NGC\,4631, and NGC\,5236) a value of 0. We further do not distinguish between bulges and pseudo-bulges.

All properties used in this work are compiled in Table \ref{tab:sample}, as are the corresponding references (to our best knowledge).

\begin{table*}[ht]
\caption{The ELVES satellite systems.}             
\centering                          
\begin{tabular}{l l r l r r l r}        
\hline\hline                 
Name & alt. name & r$_{cover}$ &  N &  N$_{250}$ &  N$_{250}^{\rm cor}$ &  B/T &  log M$_\odot$ \\    
  \\    
\hline      \\[-2mm]                  
NGC\,224 & M\,31 &     300 &   20 (a)&       17 &           17 & 0.32$\pm$0.11 (1) &       11.01 \\
 NGC\,253 & Scl &     300 &    6 (b)&        4 &            4 & 0.28$\pm$0.14 (1) &       10.77 \\
 NGC\,628 & M\,74 &     300 &   14 (b)&       11 &           11 & 0.10$\pm$0.03 (2)&       10.45 \\
 NGC\,891 & &     200 &    7 (b)&        7 &           10 & 0.20$\pm$0.05 (1) &       10.84 \\
NGC\,1023 & &     200 &   17 (b)&       17 &           20 & 0.40$\pm$0.12 (2)&       10.60 \\
NGC\,1291 & &     300 &   18 (b)&       17 &           17 & 0.57$\pm$0.17 (3)&       10.78 \\
NGC\,1808 & &     300 &   14 (b)&       13 &           13 & 0.05$\pm$0.01 (4)&       10.01 \\
NGC\,2683 & &     300 &   10 (b)(c)&        9 &            9 & 0.32$\pm$0.01 (5)&       10.50 \\
NGC\,2903 & &     300 &    7 (b)&        6 &            6 & 0.07$\pm$0.03 (1) &       10.67 \\
NGC\,3031 &M\,81 &     300 &   24 (d) &       23 &           23 & 0.46$\pm$0.15 (1) &       10.66 \\
NGC\,3115 & &     300 &   19 (b)&       17 &           17 & 0.80$\pm$0.1 (1) &       10.76 \\
NGC\,3344 & &     300 &    7 (b)&        7 &            7 & 0.01$\pm$0.01 (6)&       10.27 \\
NGC\,3379 & &     370 &   66 (b)&       40 &           40 & 1.00$\pm$0.01 &       10.63 \\
NGC\,3521 & &     330 &   12 (b)&        9 &            9 & 0.15$\pm$0.03 (7)&       10.83 \\
NGC\,3556 & M\,108 &     300 &   14 (b)&       12 &           12 & 0.00$\pm$0.01 &        9.94 \\
NGC\,3627 & M\,66 &     300 &   32 (b)&       26 &           26 & 0.14$\pm$0.02 (7)&       10.66 \\
NGC\,4258 & M\,106 &     150 &    8 (b)&        8 &           13 & 0.12$\pm$0.03 (1) &       10.62 \\
NGC\,4517 & &     300 &    9 (b)&        5 &            5 & 0.02$\pm$0.02 (8)&        9.93 \\
NGC\,4565 & &     150 &    9 (b)&        9 &           14 & 0.25$\pm$0.05 (1) &       10.88 \\
   NGC\,4594 & M\,104 &     150 &   15 (b)&       15 &           20 & 0.76$\pm$0.02 (9)&       11.09 \\
NGC\,4631 & Whale &     200 &   13 (b)(c)&       13 &           16 & 0.00$\pm$0.01 &       10.05 \\
NGC\,4736 & M\,94 &     300 &   14 (b)(e)&       12 &           12 & 0.23$\pm$0.01 (7)&       10.29 \\
NGC\,4826 & M\,64 &     300 &    9 (b)&        7 &            7 & 0.11$\pm$0.03 (2)&       10.36 \\
NGC\,5055 & M\,63 &     300 &   14 (b)&       11 &           11 & 0.18$\pm$0.02 (7)&       10.72 \\
NGC\,5128 & Cen\,A &     200 &   22 (f) &       22 &           25 & 1.00$\pm$0.01 (1) &       10.92 \\
NGC\,5194 & M\,51 &     150 &    3 (b)&        3 &            8 & 0.08$\pm$0.01 (7)&       10.73 \\
NGC\,5236 & M\,83 &     300 &   11 (g)&       10 &           10 & 0.00$\pm$0.01 &       10.37 \\
NGC\,5457 & M\,101 &     300 &    9 (b)(c)(h)&        9 &            9 & 0.01$\pm$0.01 (7)&       10.33 \\
NGC\,6744 & &     200 &   11 (b)&       11 &           14 & 0.23$\pm$0.02 (7)&       10.64 \\
\hline
\end{tabular}
\tablefoot{ The references for the dwarf galaxies are: (a) \citet{2012AJ....144....4M,2013ApJ...776...80M}, (b) \citet{2022ApJ...933...47C}, (c) \citet{2016A&A...588A..89J}, (d) \citet{2009AJ....137.3009C,2013AJ....146..126C}, (e) \citet{2018ApJ...863..152S}, (f) \citet{2014ApJ...795L..35C,2016ApJ...823...19C,2019ApJ...872...80C}, \citet{2017A&A...597A...7M,MuellerTRGB2019}, (g) \citet{2015A&A...583A..79M,2017MNRAS.465.5026C,MuellerTRGB2018}, (h) \citet{2014ApJ...787L..37M,2017ApJ...837..136D,Muller2017,2019ApJ...885..153B}. The references for bulge and total mass estimations are: (1) \citet{2017ApJ...837L...8B}, (2) \citet{2001A&A...368...16M}, (3) \citet{1975ApJS...29..193D}, (4) \citet{2021A&A...656A..60A}, (5) \citet{2016A&A...586A..98V}, (6) \citet{2000A&A...356..827V}, (7) \citet{2009ApJ...697..630F}, (8) \citet{2003MNRAS.343..665G}, and (9) \citet{2006ApJ...645..134B,2011ApJ...739...21J,2012MNRAS.423..877G}. For NGC\,3379 we set B/T to 1 because it is an elliptical galaxy. For NGC\,3556, NGC\,4631, and NGC\,5236 we set B/T to 0 because no bulges are visible.  }
\label{tab:sample}
\end{table*}

\section{The B/T ratio to the number of dwarf satellites relation}
\label{relation}

\begin{figure}[ht]
    \centering
    \includegraphics[width=\linewidth]{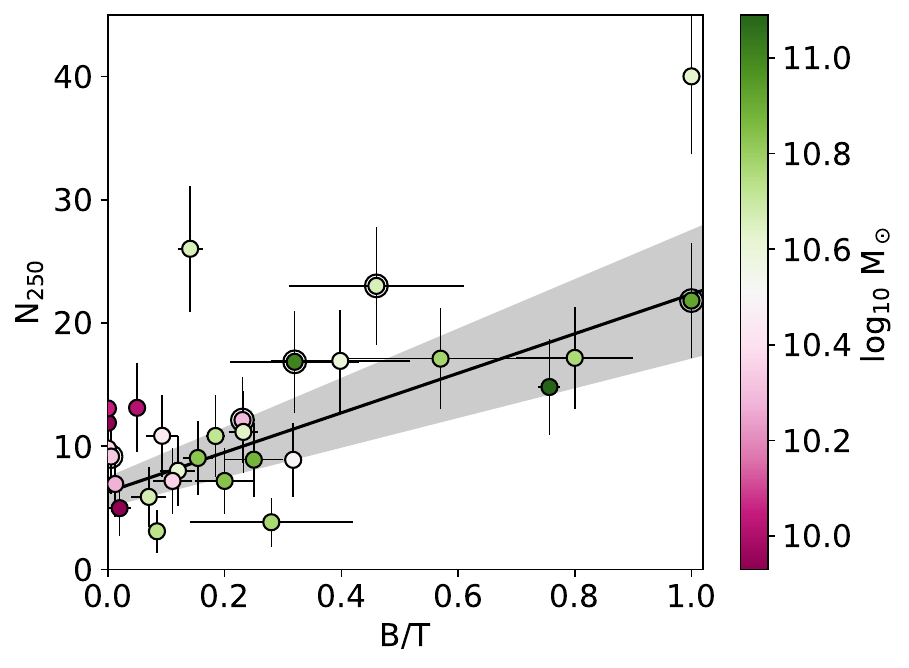}
    \caption{The number of {observed} satellites as a function of the bulge to total stellar mass relation (B/T). The dots correspond to the galaxy satellite systems, with their color representing the host stellar mass. The black line corresponds to the linear regression, the gray area to the 1$\sigma$ uncertainty.}
    \label{fig:ratios}
\end{figure}

In Fig.\,\ref{fig:ratios}  we show the number of satellites N$_{250}$ as a function of B/T. The data is color coded according to the stellar mass of the host galaxy. Looking at all data points, a positive correlation is clearly visible. A linear regression of the data finds a slope of  16.05$\pm$3.99 and an intercept of 6.29$\pm$1.15 for the uncorrected sample and 16.84$\pm$3.88 and 7.57$\pm$1.12 for the corrected sample, respectively. The errors correspond to the 1$\sigma$ uncertainty. There are almost no differences whether we consider the corrected or the uncorrected values for the two fits. Therefore, we consider that the uncorrected abundance of dwarfs N$_{250}$ represents the data well enough for the further discussion. {The same is true if we only consider the 20 galaxies with a coverage of at least 250\,kpc. The positive correlation remains with a slope of  16.72 $\pm$ 5.66 and an intercept of 6.89 $\pm$ 1.49. We conclude that the full sample represents the data well enough.}

\begin{figure}[ht]
    \centering
    \includegraphics[width=\linewidth]{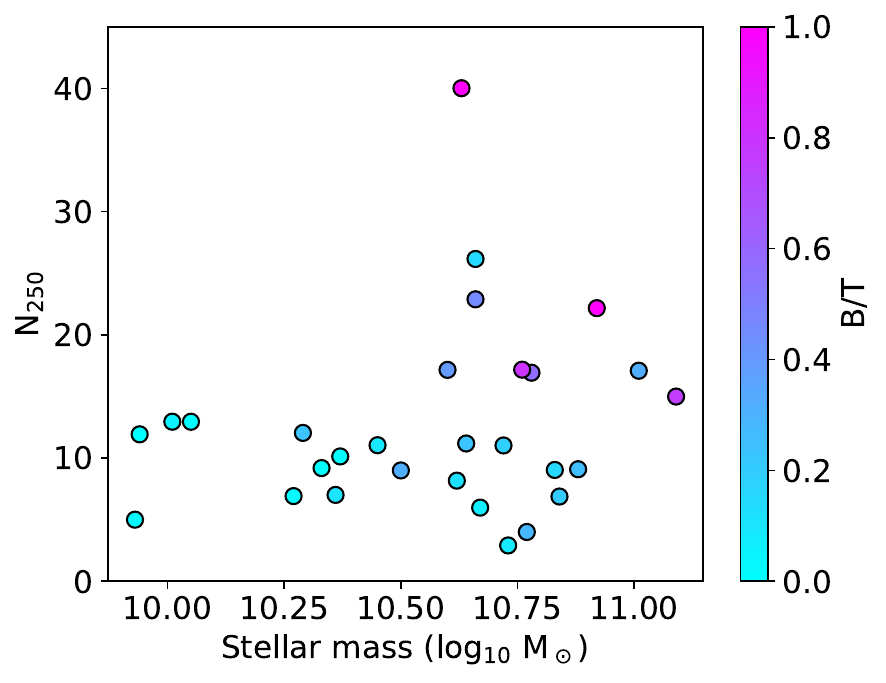}
    \caption{{The number of observed satellites as a function of the stellar mass of the host. The color coding represents the B/T ratio.} }
    \label{fig:stellar_N}
\end{figure}

Is there a dependence on the stellar mass of the host galaxy? To test that, we color code the stellar mass in Fig.\,\ref{fig:ratios} and split the sample into two with a mass cut at 10$^{10.5}$\,M$_\odot$. This split results in the massive sample with 18 hosts and light sample with 11 hosts.
For the massive sample, the slope and intercept are 21.20$\pm$5.13 and  3.89$\pm$1.83, and for the light sample it is 3.08$\pm$8.96 and  8.80$\pm$1.10. It is noteworthy that the correlation is driven by the more massive galaxies. When considering host galaxies with stellar masses below 10$^{10.5}$, the slope is consistent with zero. This may come from the fact that this sample is likely under-sampled, as all galaxies have low B/T ratios{, which is  visible in Fig.\,\ref{fig:stellar_N}, where we plot the number of satellites as a function of the stellar mass}. More data is needed there.

To estimate the significance of the correlation between B/T and N$_{250}$, we conducted a Monte Carlo simulation. For that, we ran for each of the three samples (all, larger 10$^{10.5}$\,M$_\odot$, and smaller 10$^{10.5}$\,M$_\odot$.) 10'000 iterations where we kept B/T fixed and re-assigned the number of satellites $N_{250}$ for a given giant galaxy. For each iteration, we re-evaluate the slope and intercept of the linear regression. The resulting slope and intercept distributions can be approximated with a normal distribution. Comparing the observed slope with these simulations, we measure standard deviations of 4.6$\sigma$, 3.6$\sigma$ and 0.4$\sigma$, respectively. This means that for the total sample and the massive sample, we find that the observed slope is significant ($>$3.0$\sigma$), that is, there is a positive linear relation between B/T and $N_{250}$. For the light sample, the observed slope is consistent with zero ($<$1.0$\sigma$).

Another way to check whether the relation is significant is to calculate the Pearson correlation. Calculating this for the three samples, we get Pearson $r$'s of 0.69, 0.69, and 0.02. A sample is perfectly correlated (or anti-correlated) if Pearson's $r$ value is either -1 or 1 and uncorrelated with a value of 0. Generally, a Pearson value larger than 0.5 is interpreted as highly correlated and a value between 0.0 and 0.3 as mildly correlated. With observed values of 0.7 the full and massive samples are both highly correlated. {With a value of 0.0} the light sample is uncorrelated. This gives a similar picture as before: the more massive galaxies follow a linear relation between the bulge to total stellar mass ratio  and the number of dwarf galaxies associated to the system, while the less massive galaxies do not indicate a trend. Again, for the latter this could be due to a not well-sampled sample. It will be important to test more such galaxies with a large B/T ratio to see whether this persists. For the more massive galaxies, the picture looks more robust with a better overall sampling of B/T ratios.

\begin{figure}[ht]
    \centering
    \includegraphics[width=\linewidth]{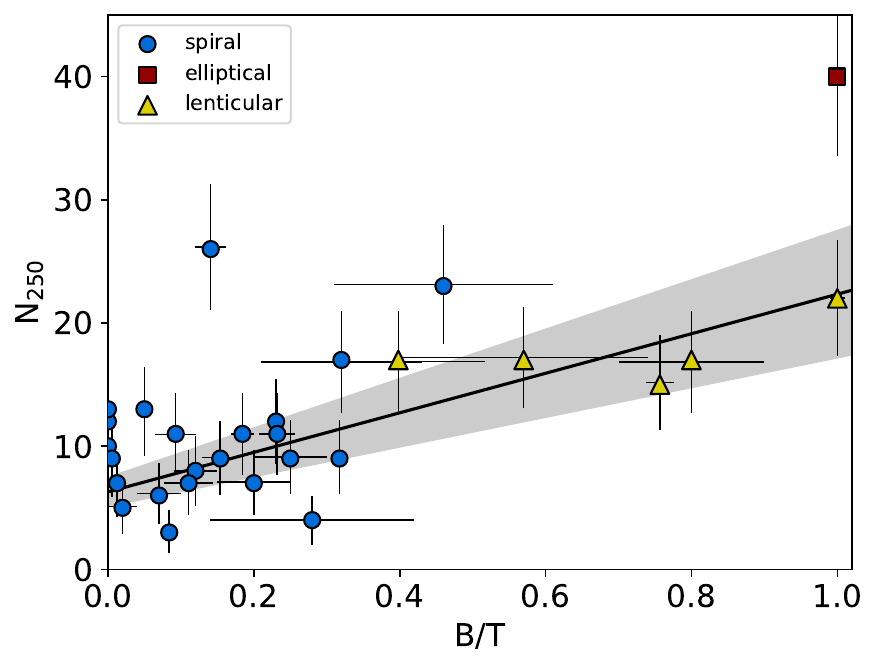}
    \caption{Same as Fig.\,\ref{fig:ratios} but color coded according to morphology. The blue dots represent spiral galaxies, the orange triangle lenticular galaxies, and the red square the elliptical galaxy in our sample. }
    \label{fig:ratios_morph}
\end{figure}

Is this relation due to morphology? Five of the galaxies (NGC\,1023, NGC\,1291, M104, NGC\,3115, and Cen\,A) in our sample are lenticular galaxies, and one  is an elliptical galaxy (NGC\,3379). Together, lenticular and elliptical galaxies are considered as early type galaxies (ETGs). The ETGs in our sample populate, unsurprisingly, the high end of the B/T ratios. If we exclude these galaxies, i.e. keep only the late type galaxies (LTGs) and repeat the previous analysis, we get a slope of 8.51$\pm$8.41 and intercept of 7.14$\pm$1.45, which is significant at a 1.2$\sigma$ level based on the Monte Carlo runs. Being below 2$\sigma$, the LTG sample by itself is not significant, meaning that we do not find a clear relation between B/T and the number of satellites. This is obvious when considering the LTGs in Fig\,\ref{fig:ratios_morph}.
Similarly, the Pearson correlation is 0.34, showing only a mild correlation. We note that the LTGs are not well sampled along the B/T ratios, with only one galaxy having a B/T ratio larger than 0.4. This galaxy -- M81 -- however, has a larger abundance of dwarfs than the average of other LTGs with lower B/T ratios. The results somewhat {increase} when considering only the LTGs with stellar masses above 10$^{10.5}$\,M$_\odot$ to a significance of 1.8$\sigma$ and a Pearson correlation of 0.48, which is however still below the common detection threshold of 3$\sigma$. Because lenticular galaxies were former spiral galaxies which have undergone morphological transformation through some mechanism such as quenching \citep{1998ApJ...495..139M,2019ApJ...878...69L}, mergers \citep{2014A&A...570A.103B}, or a more complex multi-stage formation process \citep{2021MNRAS.504.2146B}, it is not clear what a fair sample selection would be. By including the lenticular galaxies, the relation between B/T and N$_{250}$ is significant, by excluding them, it is not. This indicates that they are one of the drivers of the relation.

\section{Comparison to simulations}
\label{simulations}
The comparison of the observed B/T to number of satellites relation to simulation  was previously conducted  with dark matter-only simulations \citep{2019ApJ...870...50J}, namely the Millennium-II simulation \citep{2009MNRAS.398.1150B}, where dwarfs galaxies, bulges, and disks were "painted" to the dark matter halos based on semi-analytical models \citep{2011MNRAS.413..101G}. These semi-analytical models may not represent realistic galaxies in a $\Lambda$CDM universe. Therefore, we re-analyse the relation in a more modern cosmological simulation. 

We extracted simulated galaxies from the Illustris TNG100-1 simulation \citep{2019ComAC...6....2N}, a gravo-magnetohydrodynamical model simulating the growth of galaxies within a box of 106.5\,Mpc side length, from time z$\sim$20 to z=0 {with baryonic mass resolution of $1.4\times10^{6}M_{\odot}$ and dark matter mass resolution of $7.5\times10^{6}M_{\odot}$}. From this simulation, we select central host galaxies, defined as those halos considered to be the centre of a galaxy group, with stellar masses between $9.9<\log_{10} M_{\odot}<11.1$ at z=0. This range corresponds to the minimal and maximal stellar mass of the centrals in the ELVES catalog, respectively. {To remove highly populated galaxy clusters, we then select only galaxy groups which consist of at most 3 galaxies with absolute magnitude $M_g<-19.5$. To remove dark halos, only sub-halos with absolute magnitude $M_g<-9$ are selected. Finally, from these groups we select only sub-halos within 250$\,$kpc of the central host galaxy}. We then use the supplementary catalogue provided by \citet{2015ApJ...804L..40G} to determine the bulge to stellar mass ratio for each central galaxy, where this ratio is the mass fraction of stars that have circularity parameter $\epsilon<0$ multiplied by 2. {One caveat with this catalog that each galaxy does contain a bulge. The smallest B/T ratio in the selected Illustris-TNG100 analogs is 0.08, while the observed catalog contains galaxies with B/T ratios equal to zero.  }

 \begin{figure*}[ht]
    \centering
    \includegraphics[width=0.32\linewidth]{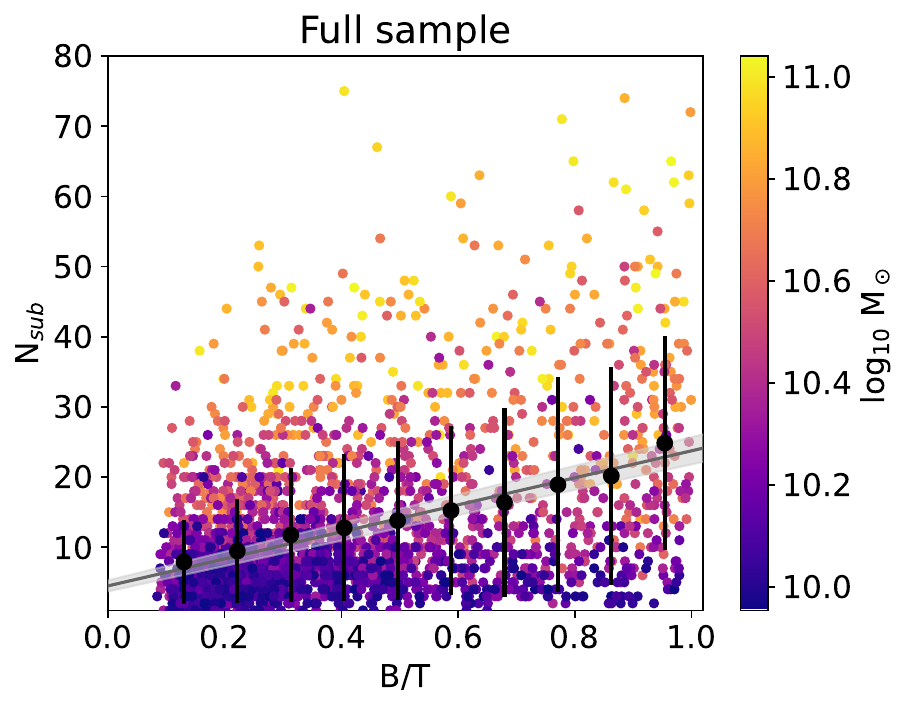}
    \includegraphics[width=0.32\linewidth]{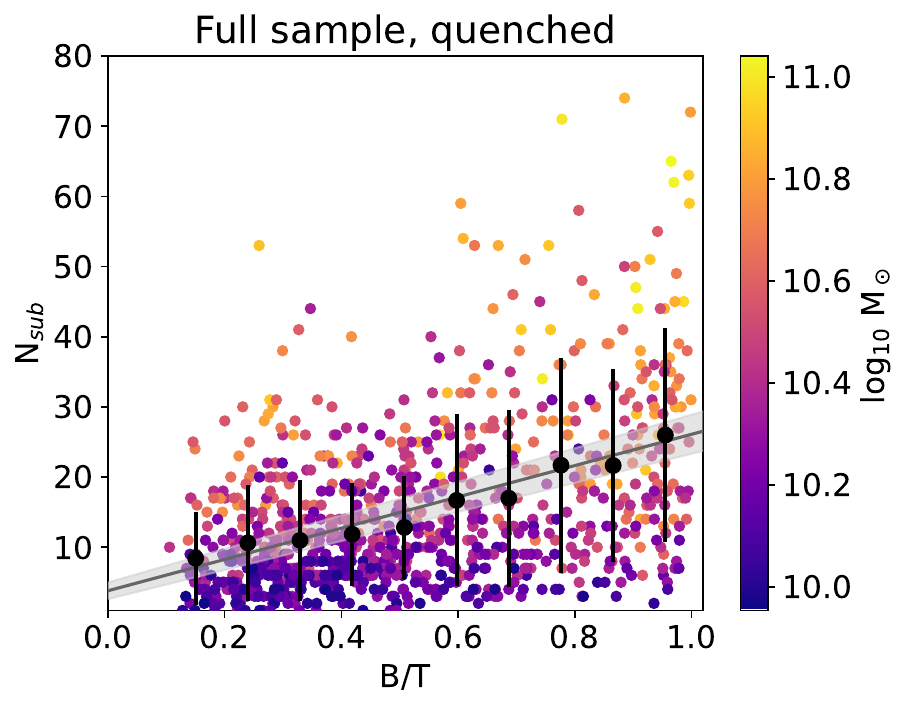}
    \includegraphics[width=0.32\linewidth]{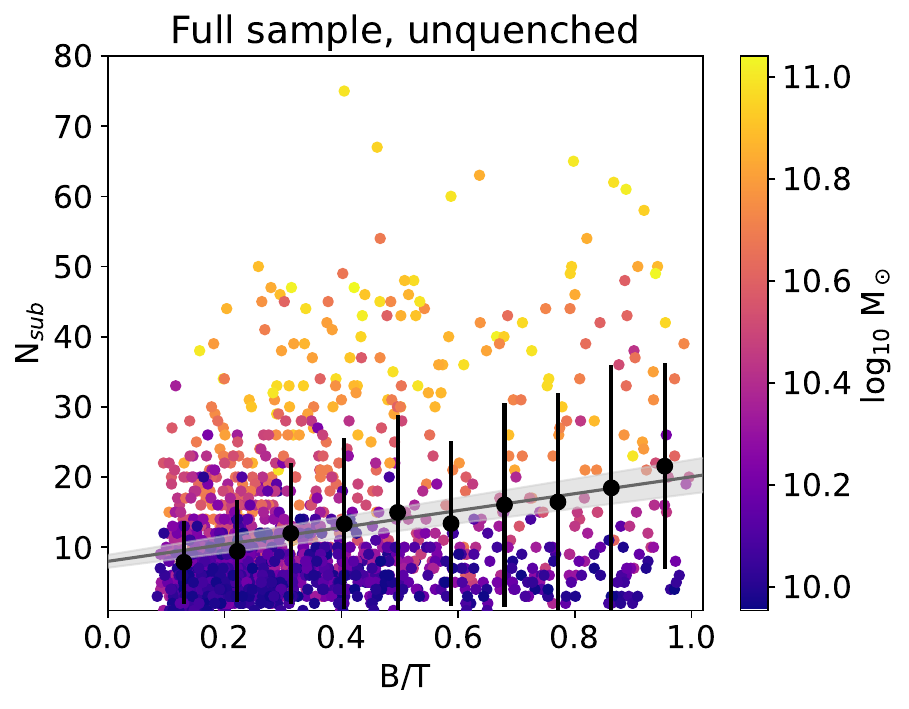}\\
    \includegraphics[width=0.32\linewidth]{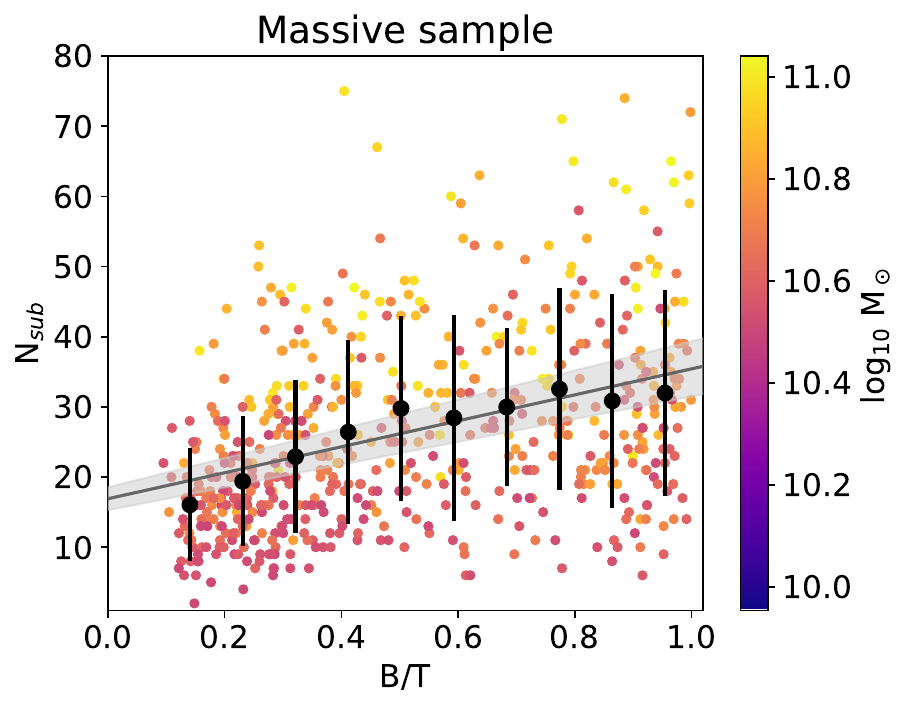}
    \includegraphics[width=0.32\linewidth]{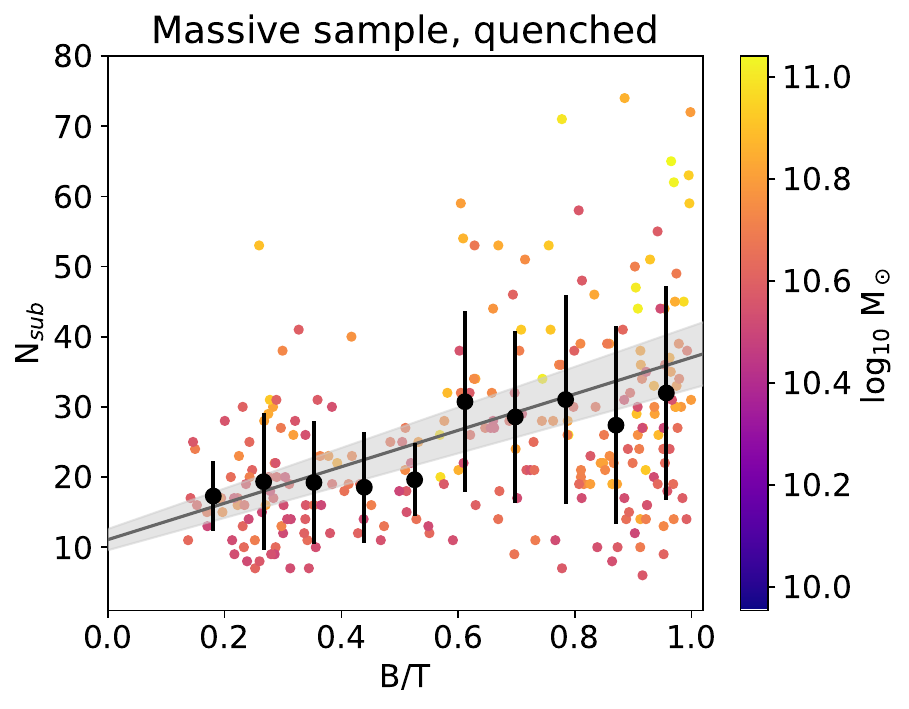}
    \includegraphics[width=0.32\linewidth]{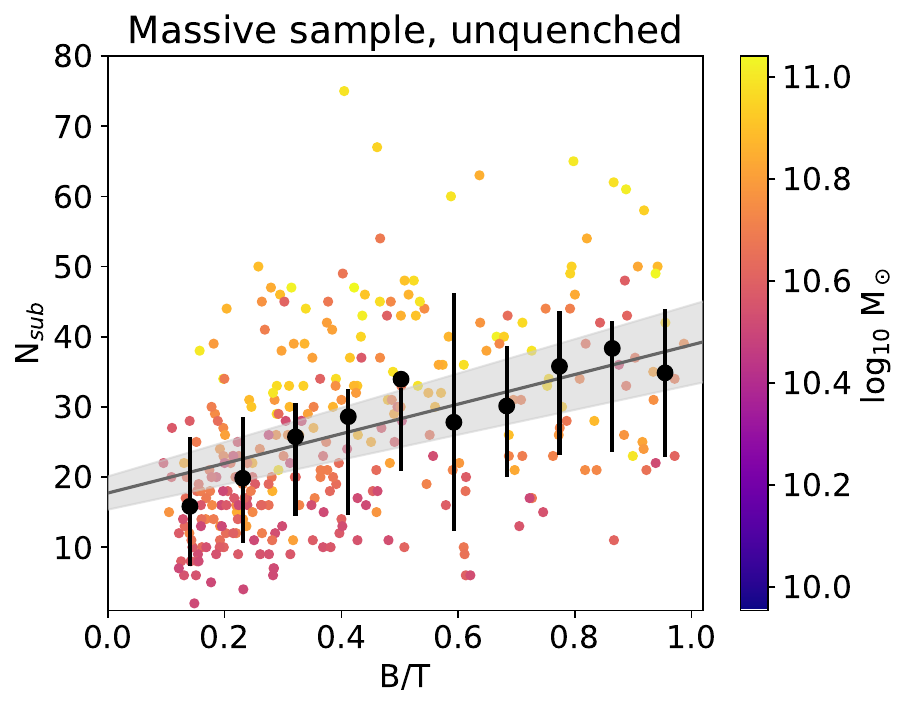}\\
    \includegraphics[width=0.32\linewidth]{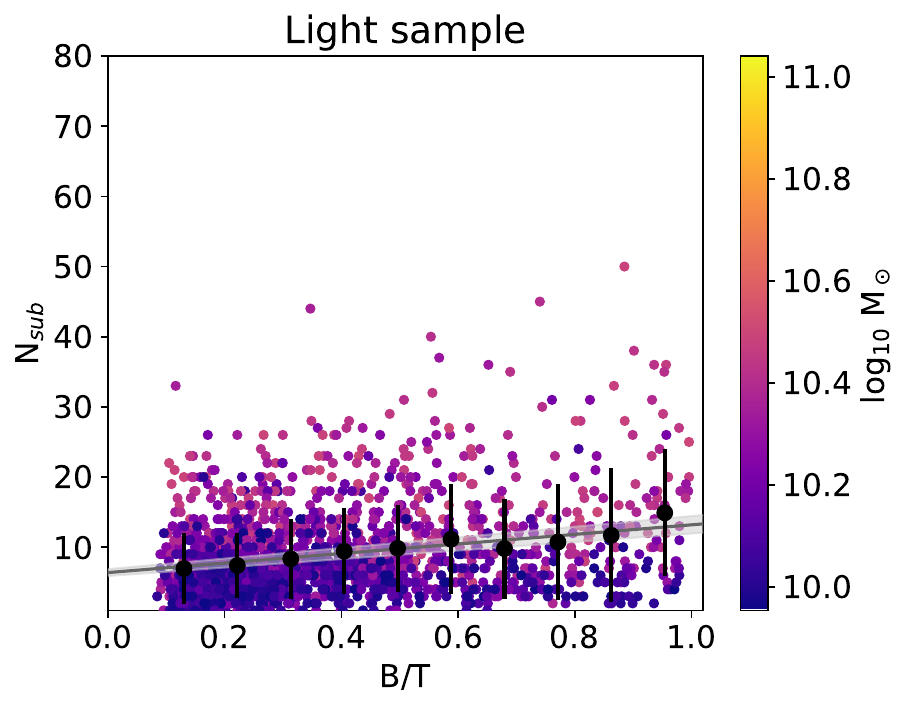}
    \includegraphics[width=0.32\linewidth]{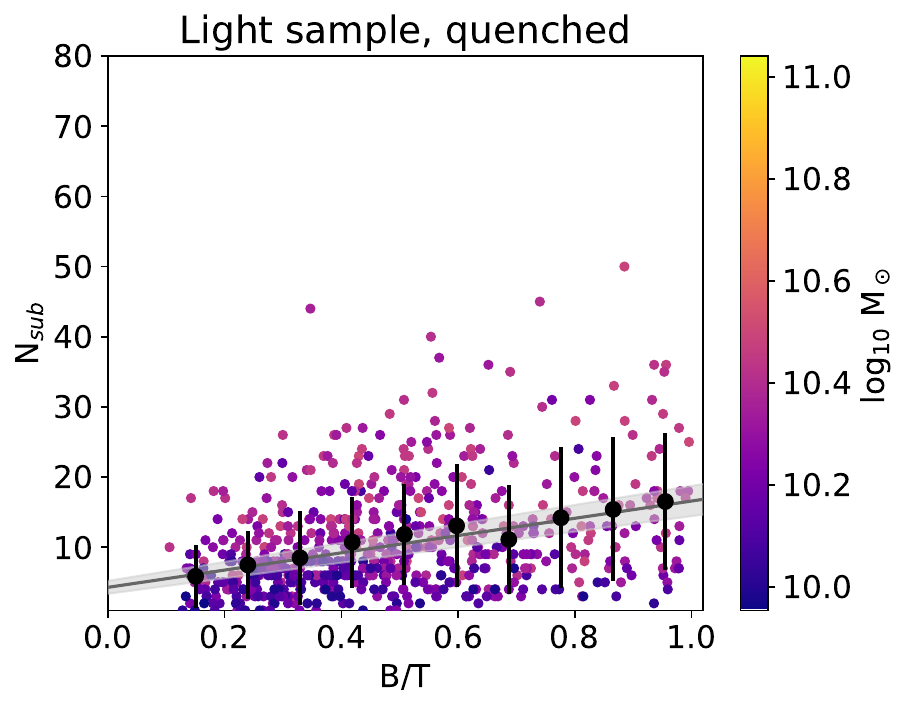}
    \includegraphics[width=0.32\linewidth]{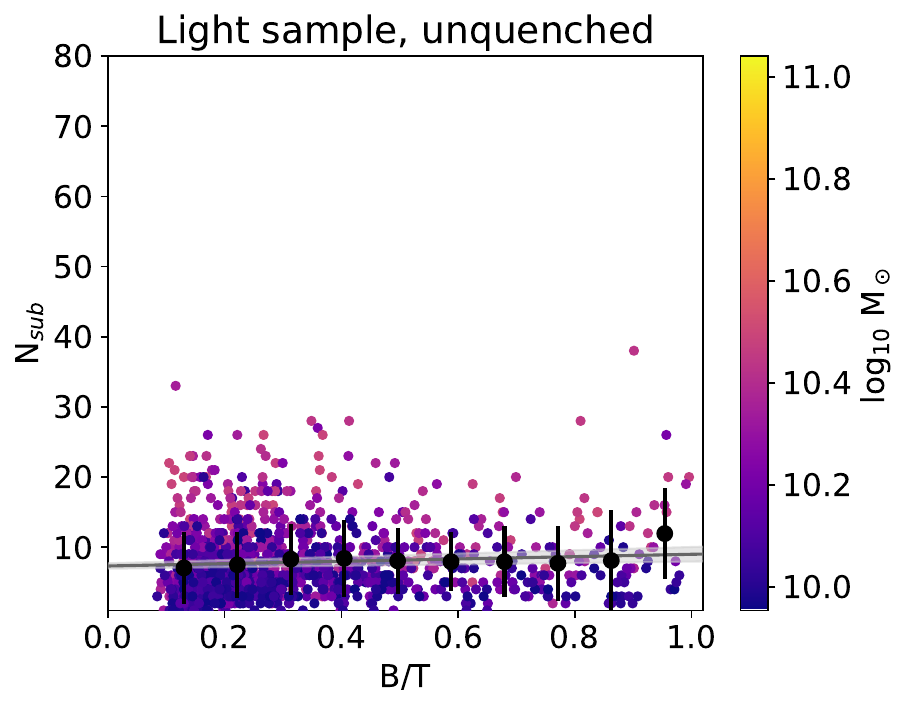}
    \caption{The number of subhalos as a function of the bulge to total stellar mass relation (B/T) in Illustris-TNG100. The colored dots correspond to the central halos and their subhalo systems, with their color representing the central halo's stellar mass. {To demonstrate the distribution of the data for a given B/T, we use a binning scheme. The black dots and uncertainties give the mean in these bins and the lines the 1$\sigma$ standard deviation within these bins. The gray line corresponds to the linear regression to all the Illustris-TNG100 halos in our sample, and the gray area to the 1$\sigma$ uncertainty.} Top: the full sample, middle: the massive sample, and bottom: the light sample. Left: the corresponding sample, middle: the quenched sample, and right: the unquenched sample.}
    \label{fig:ratios_illustris}
\end{figure*}

\begin{figure}[ht]
    \centering
    \includegraphics[width=\linewidth]{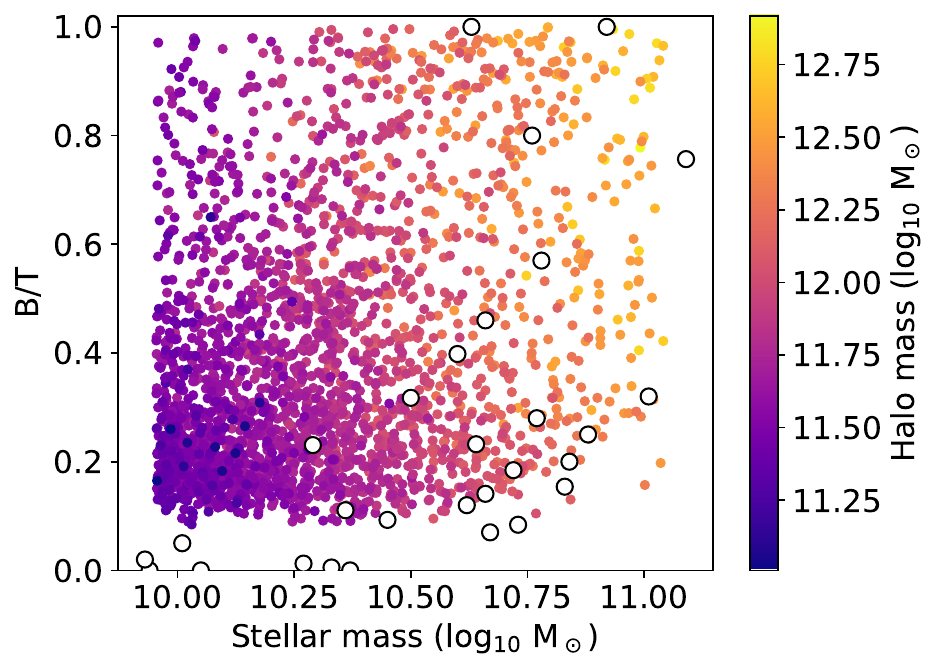}
    \caption{{The stellar mass to B/T distribution of Illustris-TNG100 galaxies and the observed groups. Colorcoded dots are the Illustris-TNG100 galaxies, with the color representing their halo mass.  The white dots represent the ELVES galaxies. }}
    \label{fig:stellar_bt}
\end{figure}

Fig.\,\ref{fig:ratios_illustris} plots the number of sub-halos contained within each of the Illustris-TNG100 analogs against the bulge mass ratio of the central halos of those analogs. Because we found that morphology drives the observed relation, we furthermore show quenched and unquenched central galaxies separately. We use a star formation rate threshold of 0.1\,M$_\odot$ per year to separate between quenched and unquenched halos, representing the ETGs and LTGs, respectively. And finally, we again make the split between the massive ($>$10$^{10.5}$\,M$_\odot$) and light ($<$10$^{10.5}$\,M$_\odot$) sample.
Contrary to previous studies based on the Millennium-II simulation, we do find a positive trend between the number of subhalos and the B/T ratio. 
 { We fit linear functions to these samples using a density-based weighting for imbalanced regression \citep{steininger2021density}. This weighting is necessary because the data is unevenly distributed}. Repeating the significance tests, we find the slopes to be significant well above the 3$\sigma$ level except for one sample at 2.6$\sigma$. 
 Only for unquenched central halos with a stellar mass $<$10$^{10.5}$\,M$_\odot$ we do not find a significant trend. {This is also visible in Fig.\,\ref{fig:ratios_illustris}, where all samples except this one show a significant slope.} We note that galaxies with stellar masses below  10$^{10.5}$\,M$_\odot$ are not well represented in our observed sample {with respect to their B/T distribution}. Therefore, for the purpose of comparing observations and simulations, {we deem} the massive sample with $>$10$^{10.5}$\,M$_\odot$ in stellar mass more appropriate. {This also becomes apparent when we compare the B/T distribution of the Illustris-TNG100 analogs (see Fig.\,\ref{fig:stellar_bt}), where for stellar masses below 10$^{10.5}$\,M$_\odot$ the B/T ratios of the observed galaxies is not well sampled, with B/T ratios close to zero. For analogs more massive than  10$^{10.5}$\,M$_\odot$, the observed galaxies span the same range as the Illustris-TNG100 galaxies.} 
 
 To conclude from this analysis, we do find a significant positive correlation between the number of subhalos and the ratio between the bulge to total stellar mass in Illustris-TNG100.

\section{Summary and conclusions}
In the past, there were suggestions that there exists a correlation between the number of dwarf galaxy satellites and the size of the bulge of the central galaxies, which may be inconsistent with $\Lambda$CDM cosmology. In this work, we tested this with a complete catalog of satellite systems within the nearby universe. Using 29 galaxy groups, we find that there indeed exists a linear relation at the 4$\sigma$ level between the number of satellites and the bulge to total stellar mass of the host, as previously suggested by \citet{2020MNRAS.493L..44J} based on a total of seven systems. We did not use two of their systems, namely the Milky Way due to a different dwarf galaxy detection methodology as the rest of our sample, and M\,33, which has a lower stellar mass than what we consider here, so our sample can be considered as an -- almost -- different data set. Because we used in our analysis host galaxies with similar stellar masses, the signal is not driven by the well known relation between the halo mass and the number of satellites. 

Is the B/T to number of satellites relation perhaps driven by morphology? In our sample are both early-type (6) and late-type (23) galaxies.  The ETGs -- more specifically the five lenticular galaxies -- seem to be the drivers of the relation at the high B/T end. These galaxies have similar stellar masses as the majority of the LTGs in our sample and should therefore, naively, not host more dwarf galaxies. Excluding these ETGs from our sample, the B/T to number of satellites relation vanishes, however, the B/T range is not well sampled, with only one LTG having a B/T ratio larger than 0.4. This LTG -- M\,81 -- hosts more satellites than the rest of LTGs. Is this just a coincidence?

{Lenticular galaxies are intermediates between spirals and ellipticals and undergo a morphological transformation. It is therefore not clear whether removing them from the sample is fair or even useful. Assuming the relation between the  B/T ratio and number of satellites is real, their morphology may give us clues about the origin of the relation. {One speculative idea is that major mergers may be responsible for the large number of satellites as well as the build up of the central bulge.}
The lenticular galaxy Cen\,A underwent a recent major merger 2\,Gyr ago with a mass ratio of its progenitors of up to 1.5 \citep{2020MNRAS.498.2766W}. Cen\,A has a stellar mass of 8$\times$10$^{10}$\,M$_\odot$ and 22 known satellites within $N_{250}$. By assuming that both progenitors have stellar masses between 3 to 5$\times$10$^{10}$\,M$_\odot$ we can guess how many dwarf galaxies each of them would bring with them.  Taking the ELVES catalog as reference, eight central galaxies are in this range of stellar masses with a median of 14 dwarf galaxies, so the combined satellite population of two of these progenitor galaxies (i.e. 28 dwarf galaxies) would be consistent with the number of observed satellites around Cen\,A ($N_{250}$=22). Of course, this simplified assessment does not take into account any mass loss {or star formation} during the merger, as well as any disruption of dwarf galaxies, but may still point towards an explanation of this empirical relation. The M\,81 system provides further arguments for this interpretation. For M\,81  there is evidence for ongoing gas stripping with NG\,3077 \citep{1994Natur.372..530Y}, a dwarf elliptical on the more massive end. M\,81 is transforming towards a lenticular galaxy, with a sluggish star formation rate (0.25 M$_\odot$/yr, \citealt{2019ApJS..243....3L}). The M\,81 group also possesses M\,82 within its virial radius. M\,82 is a starburst galaxy with stellar mass 10$^{10.5}$\,M$_\odot$. M\,81 could have acquired satellites from M\,82, boosting its abundance of dwarf galaxies, as we hypothesised for Cen\,A. Ultimately, however, this scenario would need to be quantified by simulations of mergers.}

Because the catalog is limited by a 10\,Mpc cut, there is not much room for improvements on the data catalog. Therefore, to increase the sample we need to go out to galaxies further away. Within 10-40\,Mpc, deep targeted observations are capable of detecting dwarf galaxies to similar limits as discussed here, and especially wide surveys like Euclid or LSST will increase the sample by a manifold. This will help to better sample the different B/T ratios, especially at larger values. It will be particularly interesting to sample spiral galaxies with high B/T values to see whether the observed relation is driven by morphology or not. 

Previously, it was found that a relation between a the size of the bulge of the central galaxies and its satellite population is not expected in standard cosmology  \citep{2019ApJ...870...50J}. However, \citet{2019ApJ...870...50J} used the  Millennium-II simulation, which is a dark matter-only simulation. Because this simulation does not include baryonic physics, semi-analytical models were applied to "paint" bulges and disks to dark matter halos, which may not represent realistic galaxies.
Here, we instead use the Illustris-TNG100 simulation which includes baryonic physics to re-evaluate the findings from \citet{2019ApJ...870...50J}. Contrary to the findings from the Millennium-II simulation, there is a correlation between the bulge to disk ratio and the number of subhalos/dwarf galaxy satellites. This indicates that $\Lambda$CDM generally does produce something like the observed relation. We investigated if there is a dependence on the stellar mass of the host galaxy, as well the star formation rate. By splitting the Illustris-TNG100 analogs into a massive and light sample, as well as a quenched and unquenched sample, we investigated the appearance of a correlation between the bulge to disk ratio and the number of subhalos. For all but the unquenched sample with stellar mass below 10$^{10.5}$\,M$_\odot$ we do find a significant positive trend. For the unquenched sample in this mass range the Pearson correlation suggests a weak correlation, however, fitting a linear function gives a slope consistent with zero. This sample, however, is not well represented in the ELVES catalog. 

{As a caveat to the analysis we want to point out that in $\Lambda$CDM the number of subhalos is driven by the halo mass, which is a property that we have access to in simulations, but is difficult to measure in galaxy groups, with large uncertainties for even well studied systems \citep[e.g., ][]{2005AJ....129..178K,2022A&A...662A..57M}. As a proxy, we used the stellar mass of the host galaxy, which is related to the halo mass by the stellar-to-halo mass relation (SHMR). The SHMR has a scatter of roughly 0.2\,dex \citep[e.g., ][]{2013ApJ...771...30R,2018AstL...44....8K} which we did not consider here. For future work, a systematic study of the halo masses of the ELVES target could bring new insights into our analysis.}

In general, Illustris-TNG100 seems to produce a positive correlation between the number of satellites and the bulge to total stellar mass ratio.  We conclude that {the observered empirical} relation {may not be} in tension with standard $\Lambda$CDM cosmology {after all}.

\label{summary}

\begin{acknowledgements} 
{We thank the referee for the constructive report, which helped to clarify and improve the manuscript.}
O.M. thanks Eva Schnider for helpful discussions on the fitting and visualization of the Illustris-TNG100 data. {O.M. also thanks Benoit Famaey for discussions on the origin of dwarf galaxies in MOND.}
O.M. is grateful to the Swiss National Science Foundation for financial support under the grant number PZ00P2\_202104. 
\end{acknowledgements}

\bibliographystyle{aa}
\bibliography{bibliographie}

\end{document}